\documentclass[table,10pt]{interact}
\usepackage[T1]{fontenc}
\usepackage{times}
	\usepackage[utf8]{inputenc}
	\usepackage{csquotes}
\usepackage{graphicx}
\usepackage[svgnames]{xcolor}
\usepackage{listingsutf8}
\usepackage{multirow}
	\MakeInnerQuote{"}
	\usepackage{eurosym} 
\usepackage{tikz,amssymb,graphicx,fontawesome}
\usetikzlibrary{fit}

\usepackage{longtable}
\usepackage{changes}

\usepackage[natbibapa,nodoi]{apacite}
	\usepackage{textcomp}
	\usepackage{amsmath,amsfonts}
	\usepackage{multido}

	\usepackage{hyperref}
	\hypersetup{%
		 colorlinks = {true},
		 urlcolor = {blue},
		 linkcolor = {black},
		 citecolor = {black},
		 pdfauthor = {Martial Phélippé-Guinvarc'h},
		 pdftitle = {French Wine Revenue Map   },
		 pdfkeywords = {crop yield insurance, vineyard map}
	}

\DeclareUnicodeCharacter{20AC}{\EUR{}}

\usepackage{listingsutf8}
\definecolor{mbfaulmbleu}{RGB}{15,91,164}
\definecolor{mbfaulmbleumarine}{RGB}{40,56,120}
\definecolor{mbfaulmorangesec}{RGB}{228,82,65}
\definecolor{mbfaulmorange}{RGB}{228,66,45}
\definecolor{mbfaulmgrissec}{RGB}{91,93,95}
\definecolor{mbfaulmgris}{RGB}{112,111,111}
\definecolor{mbfaulmcaption}{RGB}{228,66,45}
\definecolor{DEG}{RGB}{228,0,56}
\definecolor{IRAorange}{RGB}{229,68,46}
\definecolor{IRAbleu}{RGB}{40,57,121}

\lstdefinestyle{mbfaulmR} {
	language=R,
	inputencoding=utf8/latin1,
	backgroundcolor=\color{mbfaulmgrissec!5},   
	frame=lines,
	basicstyle=\footnotesize\ttfamily, 
	keywordstyle = {\bfseries\color{mbfaulmbleumarine}\scshape},
	identifierstyle=\color{mbfaulmbleu},
	stringstyle ={\ttfamily\color{mbfaulmorangesec}},
	commentstyle=\color{gray}, 
	showstringspaces=false,
	numbers=left, numberstyle=\tiny, stepnumber=1, numbersep=5pt, %
	rulesepcolor=\color{black},
	columns=fullflexible,
	breaklines=true,
	breakatwhitespace=false,
	emph={[1] (,),[,]}, emphstyle={[1] \color{mbfaulmorange}},
	emph={[2] {,}}, emphstyle={[2] \color{green}}
	literate=
	{á}{{\'a}}1 {é}{{\'e}}1 {í}{{\'i}}1 {ó}{{\'o}}1 {ú}{{\'u}}1
	{À}{{\`A}}1 {à}{{\`a}}1
	{Á}{{\'A}}1 {É}{{\'E}}1 {Í}{{\'I}}1 {Ó}{{\'O}}1 {Ú}{{\'U}}1
	{à}{{\`a}}1 {è}{{\`e}}1 {ì}{{\`i}}1 {ò}{{\`o}}1 {ù}{{\`u}}1
	{À}{{\`A}}1 {È}{{\'E}}1 {Ì}{{\`I}}1 {Ò}{{\`O}}1 {Ù}{{\`U}}1
	{ä}{{\"a}}1 {ë}{{\"e}}1 {ï}{{\"i}}1 {ö}{{\"o}}1 {ü}{{\"u}}1
	{Ä}{{\"A}}1 {Ë}{{\"E}}1 {Ï}{{\"I}}1 {Ö}{{\"O}}1 {Ü}{{\"U}}1
	{â}{{\^a}}1 {ê}{{\^e}}1 {î}{{\^i}}1 {ô}{{\^o}}1 {û}{{\^u}}1
	{Â}{{\^A}}1 {Ê}{{\^E}}1 {Î}{{\^I}}1 {Ô}{{\^O}}1 {Û}{{\^U}}1
	{œ}{{\oe}}1 {Œ}{{\OE}}1 {æ}{{\ae}}1 {Æ}{{\AE}}1 {ß}{{\ss}}1
	{ç}{{\c c}}1 {Ç}{{\c C}}1 {ø}{{\o}}1 {å}{{\r a}}1 {Å}{{\r A}}1
	{€}{{\EUR}}1 {£}{{\pounds}}1
}

\lstdefinelanguage{SAS}{
	sensitive=true,
	alsoletter={\%\&},
	keywordsprefix=\&,
	morecomment=[f][\color{DarkOliveGreen}][0]*,
	morecomment=[s][\color{DarkOliveGreen}]{***}{;},
	morecomment=[s][\color{DarkOliveGreen}]{/*}{*/},
	morecomment=[n][\color{DarkOliveGreen}]{/*}{*/},
	morecomment=[l][\color{mbfaulmgrissec!5}]{*;},
	morestring=[b]",
	morestring=[d]',
	morecomment=[s][\itshape\color{mbfaulmbleu!50}]{datalines;}{;},
	morecomment=[s][\itshape\color{mbfaulmbleu!50}]{cards;}{;},
	morekeywords=[2]{
		DATA ,PROC ,RUN ,QUIT, ODS,OPTIONS,SOLVE
	},
	morekeywords=[3]{
		\%BQUOTE ,\%NRBQUOTE ,\%CMPRES ,\%QCMPRES ,\%COMPSTOR ,\%DATATYP
		,\%DISPLAY ,\%DO ,\%ELSE ,\%END ,\%EVAL ,\%GLOBAL ,\%GOTO ,\%IF
		,\%INDEX ,\%INPUT ,\%KEYDEF ,\%LABEL ,\%LEFT ,\%LENGTH ,\%LET
		,\%LOCAL ,\%LOWCASE ,\%MACRO ,\%MEND ,\%NRBQUOTE ,\%NRQUOTE
		,\%NRSTR ,\%PUT ,\%QCMPRES ,\%QLEFT ,\%QLOWCASE ,\%QSCAN ,\%QSUBSTR
		,\%QSYSFUNC ,\%QTRIM ,\%QUOTE ,\%QUPCASE ,\%SCAN ,\%STR ,\%SUBSTR
		,\%SUPERQ ,\%SYSCALL ,\%SYSEVALF ,\%SYSEXEC ,\%SYSFUNC ,\%SYSGET
		,\%SYSLPUT ,\%SYSPROD ,\%SYSRC ,\%SYSRPUT ,\%THEN ,\%TO ,\%TRIM
		,\%UNQUOTE ,\%UNTIL ,\%UPCASE ,\%VERIFY ,\%WHILE ,\%WINDOW	},
	morekeywords=[4]{
		DO ,IF ,THEN ,ELSE ,END ,UNTIL ,WHILE
		,ABORT ,ARRAY ,ATTRIB ,BY ,CALL ,CARDS ,CARDS4 ,CATNAME ,CONTINUE
		,DATALINES ,DATALINES4 ,DELETE ,DELIM ,DELIMITER ,DISPLAY ,DM ,DROP
		,ENDSAS ,ERROR ,FILE ,FILENAME ,FOOTNOTE ,FORMAT ,GOTO ,IN ,INFILE
		,INFORMAT ,INPUT ,KEEP ,LABEL ,LEAVE ,LENGTH ,LIBNAME ,LINK ,LIST
		,LOSTCARD ,MERGE ,MISSING ,MODIFY ,OPTIONS ,OUTPUT ,OUT ,PAGE ,PUT
		,REDIRECT ,REMOVE ,RENAME ,REPLACE ,RETAIN ,RETURN ,SELECT ,SET
		,SKIP ,STARTSAS ,STOP ,TITLE ,UPDATE ,WAITSAS ,WHERE ,WINDOW ,X
		,SYSTASK,SCATTER, HBAR, VAR, VBAR, SERIES, HISTOGRAM,
		NOPRINT, PLOT=
		,ADD ,AND ,ALTER ,AS ,CASCADE ,CHECK ,CREATE ,DELETE ,DESCRIBE
		,DISTINCT ,DROP ,FOREIGN ,FROM ,GROUP ,HAVING ,INDEX ,INSERT ,INTO
		,IN ,KEY ,LIKE ,MESSAGE ,MODIFY ,MSGTYPE ,NOT ,ON ,OR ,ORDER
		,PRIMARY ,REFERENCES ,RESET ,RESTRICT ,SELECT ,SET ,TABLE ,UNIQUE
		,UPDATE ,VALIDATE ,VIEW ,WHERE,
		READ ,CON ,NUM 	},
	morekeywords=[5]{
		START,
		FINISH,
		USE,
		RETURN,
		STORE,
		DIMENSION,
		ABS	,
		ADDR	,
		ADDRLONG	,
		AIRY	,
		ALLCOMB	,
		ALLPERM	,
		ANORM420	,
		ANYALNUM	,
		ANYALPHA	,
		ANYCNTRL	,
		ANYDIGIT	,
		ANYFIRST	,
		ANYGRAPH	,
		ANYLOWER	,
		ANYNAME	,
		ANYPRINT	,
		ANYPUNCT	,
		ANYSPACE	,
		ANYUPPER	,
		ANYXDIGIT	,
		ARCOS	,
		ARCOSH	,
		ARSIN	,
		ARSINH	,
		ARTANH	,
		ATAN	,
		ATAN2	,
		ATTRC	,
		ATTRN	,
		AVG	,
		BAND	,
		BASECHAR	,
		BETA	,
		BETAINV	,
		BITFNBOR	,
		BLACKCLPRC	,
		BLACKPTPRC	,
		BLKSHCLPRC	,
		BLKSHPTPRC	,
		BLSHIFT	,
		BNOT	,
		BOR	,
		BRSHIFT	,
		BXOR	,
		BYTE	,
		CAST	,
		CAT	,
		CATQ	,
		CATS	,
		CATT	,
		CATX	,
		CDF	,
		CEIL	,
		CEILZ	,
		CEXIST	,
		CHAR	,
		CHARACTER_LENGTH	,
		CHOOSEC	,
		CHOOSEN	,
		CINV	,
		CLIBEXIST	,
		CLOSE	,
		CMISS	,
		CMP	,
		CMPT	,
		CNONCT	,
		COALESCE	,
		COALESCEC	,
		COLLATE	,
		COMB	,
		COMPARE	,
		COMPBL	,
		COMPFUZZ	,
		COMPGED	,
		COMPLEV	,
		COMPOUND	,
		COMPRESS	,
		CONSTANT	,
		CONVX	,
		CONVXP	,
		COS	,
		COSH	,
		COT	,
		COUNT	,
		COUNTC	,
		COUNTW	,
		CSC	,
		CSS	,
		CUMIPMT	,
		CUMPRINC	,
		CUROBS	,
		CURRENT_DATE	,
		CURRENT_LOCALE	,
		CURRENT_TIME	,
		CURRENT_TIME_GMT	,
		CURRENT_TIMESTAMP	,
		CURRENT_TIMESTAMP_GMT	,
		CV	,
		DACCDB	,
		DACCDBSL	,
		DACCSL	,
		DACCSYD	,
		DACCTAB	,
		DAIRY	,
		DATDIF	,
		DATE	,
		DATEJUL	,
		DATEPART	,
		DATETIME	,
		DAY	,
		DCLOSE	,
		DCREATE	,
		DEGREES	,
		DEPDB	,
		DEPDBSL	,
		DEPSL	,
		DEPSYD	,
		DEPTAB	,
		DEQUOTE	,
		DEVIANCE	,
		DHMS	,
		DIF	,
		DIGAMMA	,
		DIM	,
		DINFO	,
		DIVIDE	,
		DNUM	,
		DOPEN	,
		DOPTNAME	,
		DOPTNUM	,
		DOSUBL	,
		DREAD	,
		DROPNOTE	,
		DSNAME	,
		DSNCATLGD	,
		DUR	,
		DURP	,
		E	,
		EFFRATE	,
		ENCODCOMPAT	,
		ENCODISVALID	,
		ENVLEN	,
		ERF	,
		ERFC	,
		EUCLID	,
		EUROCURR	,
		EXIST	,
		EXP	,
		FACT	,
		FAPPEND	,
		FCLOSE	,
		FCOL	,
		FCOPY	,
		FDELETE	,
		FETCH	,
		FETCHOBS	,
		FEXIST	,
		FGET	,
		FILEEXIST	,
		FILENAME	,
		FILEREF	,
		FINANCE	,
		FIND	,
		FINDC	,
		FINDW	,
		FINFO	,
		FINV	,
		FIPNAME	,
		FIPNAMEL	,
		FIPSTATE	,
		FIRST	,
		FLOOR	,
		FLOORZ	,
		FMTINFO	,
		FNONCT	,
		FNOTE	,
		FOPEN	,
		FOPTNAME	,
		FOPTNUM	,
		FPOINT	,
		FPOS	,
		FPUT	,
		FREAD	,
		FREWIND	,
		FRLEN	,
		FROOT ,
		FSEP	,
		FUZZ	,
		FWRITE	,
		GAMINV	,
		GAMMA	,
		GARKHCLPRC	,
		GARKHPTPRC	,
		GCD	,
		GEODIST	,
		GEOMEAN	,
		GEOMEANZ	,
		GETCASURL	,
		GETLCASLIB	,
		GETLOCENV	,
		GETLSESSREF	,
		GETLTAG	,
		GETOPTION	,
		GETPXLANGUAGE	,
		GETPXLOCALE	,
		GETPXREGION	,
		GETSESSOPT	,
		GETVARC	,
		GETVARN	,
		GITFN_CLONE	,
		GITFN_CO_BRANCH	,
		GITFN_COMMIT	,
		GITFN_COMMIT_GET	,
		GITFN_COMMIT_LOG	,
		GITFN_COMMITFREE	,
		GITFN_DEL_BRANCH	,
		GITFN_DEL_REPO	,
		GITFN_DIFF	,
		GITFN_DIFF_FREE	,
		GITFN_DIFF_GET	,
		GITFN_DIFF_IDX_F	,
		GITFN_IDX_ADD	,
		GITFN_IDX_REMOVE	,
		GITFN_MRG_BRANCH	,
		GITFN_NEW_BRANCH	,
		GITFN_PULL	,
		GITFN_PUSH	,
		GITFN_RESET	,
		GITFN_RESET_FILE	,
		GITFN_STATUS	,
		GITFN_STATUS_GET	,
		GITFN_STATUSFREE	,
		GITFN_VERSION	,
		GRAYCODE	,
		GRDSVC_ENABLE	,
		GRDSVC_GETADDR	,
		GRDSVC_GETINFO	,
		GRDSVC_GETNAME	,
		GRDSVC_HOSTLIST	,
		GRDSVC_NNODES	,
		GRDSVC_OPTSETS	,
		HARMEAN	,
		HARMEANZ	,
		HASHING	,
		HASHING_FILE	,
		HASHING_HMAC	,
		HASHING_HMAC_FILE	,
		HASHING_HMAC_INIT	,
		HASHING_INIT	,
		HASHING_PART	,
		HASHING_TERM	,
		HBOUND	,
		HMS	,
		HOLIDAY	,
		HOLIDAYCK	,
		HOLIDAYCOUNT	,
		HOLIDAYNAME	,
		HOLIDAYNX	,
		HOLIDAYNY	,
		HOLIDAYTEST	,
		HOUR	,
		HTMLDECODE	,
		HTMLENCODE	,
		IBESSEL	,
		IFC	,
		IFN	,
		IFNULL	,
		INDEX	,
		INDEXC	,
		INDEXW	,
		INPUT	,
		INPUTC	,
		INPUTN	,
		INT	,
		INTCINDEX	,
		INTCK	,
		INTCYCLE	,
		INTDT	,
		INTFIT	,
		INTFMT	,
		INTGET	,
		INTINDEX	,
		INTNEST	,
		INTNX	,
		INTRR	,
		INTSEAS	,
		INTSHIFT	,
		INTTEST	,
		INTTS	,
		INTZ	,
		IORCMSG	,
		IPMT	,
		IQR	,
		IRR	,
		ISSKIPPED ,
		ISEMPTY ,
		JBESSEL	,
		JULDATE	,
		JULDATE7	,
		KCOMPARE	,
		KCOMPRESS	,
		KCOUNT	,
		KCOUNTC	,
		KCOUNTW	,
		KCOUNTX	,
		KCVT	,
		KFIND	,
		KFINDC	,
		KFINDW	,
		KINDEX	,
		KINDEXB	,
		KINDEXC	,
		KINDEXCB	,
		KLEFT	,
		KLENGTH	,
		KLOWCASE	,
		KPROPCASE	,
		KPROPCHAR	,
		KPROPDATA	,
		KREVERSE	,
		KRIGHT	,
		KSCAN	,
		KSCANX	,
		KSTRCAT	,
		KSTRIP	,
		KSUBSTR	,
		KSUBSTRB	,
		KSUBSTRN	,
		KTRANSLATE	,
		KTRIM	,
		KTRUNCATE	,
		KUPCASE	,
		KUPDATE	,
		KUPDATEB	,
		KUPDATES	,
		KURTOSIS	,
		KVERIFY	,
		KVERIFYB	,
		LAG	,
		LARGEST	,
		LBOUND	,
		LCM	,
		LCOMB	,
		LEFT	,
		LENGTH	,
		LENGTHC	,
		LENGTHM	,
		LENGTHN	,
		LEXCOMB	,
		LEXCOMBI	,
		LEXPERK	,
		LEXPERM	,
		LFACT	,
		LGAMMA	,
		LIBNAME	,
		LIBREF	,
		LOG	,
		LOG10	,
		LOG1PX	,
		LOG2	,
		LOG4SAS_APPENDER	,
		LOG4SAS_LOGEVENT	,
		LOG4SAS_LOGGER	,
		LOGBETA	,
		LOGCDF	,
		LOGISTIC	,
		LOGPDF	,
		LOGSDF	,
		LOWCASE	,
		LPERM	,
		LPNORM	,
		MAD	,
		MAKEDATE	,
		MAKETIME	,
		MAKETIMESTAMP	,
		MARGRCLPRC	,
		MARGRPTPRC	,
		MAX	,
		MCIPISLP	,
		MCIPISTR	,
		MD5	,
		MDY	,
		MEAN	,
		MEDIAN	,
		MIN	,
		MINUTE	,
		MISSING	,
		MOD	,
		MODEXIST	,
		MODULE	,
		MODULEC	,
		MODULEN	,
		MODZ	,
		MONTH	,
		MOPEN	,
		MORT	,
		MSPLINT	,
		MVALID	,
		N	,
		NDIMS	,
		NETPV	,
		NLDATE	,
		NLDATM	,
		NLITERAL	,
		NLTIME	,
		NMISS	,
		NOMRATE	,
		NORMAL	,
		NOTALNUM	,
		NOTALPHA	,
		NOTCNTRL	,
		NOTDIGIT	,
		NOTE	,
		NOTFIRST	,
		NOTGRAPH	,
		NOTLOWER	,
		NOTNAME	,
		NOTPRINT	,
		NOTPUNCT	,
		NOTSPACE	,
		NOTUPPER	,
		NOTXDIGIT	,
		NPV	,
		NULL	,
		NVALID	,
		NWKDOM	,
		OCTET_LENGTH	,
		OPEN	,
		ORDINAL	,
		PATHNAME	,
		PCTL	,
		PDF	,
		PEEK	,
		PEEKC	,
		PEEKCLONG	,
		PEEKLONG	,
		PERM	,
		PI	,
		PMT	,
		POINT	,
		POISSON	,
		POWER	,
		PPMT	,
		PROBBETA	,
		PROBBNML	,
		PROBBNRM	,
		PROBCHI	,
		PROBDF	,
		PROBF	,
		PROBGAM	,
		PROBHYPR	,
		PROBIT	,
		PROBMC	,
		PROBMED	,
		PROBNEGB	,
		PROBNORM	,
		PROBT	,
		PROPCASE	,
		PRXCHANGE	,
		PRXMATCH	,
		PRXPAREN	,
		PRXPARSE	,
		PRXPOSN	,
		PTRLONGADD	,
		PUT	,
		PUTC	,
		PUTN	,
		PVP	,
		QTR	,
		QUANTILE	,
		QUOTE	,
		RADIANS	,
		RANBIN	,
		RANCAU	,
		RAND	,
		RANEXP	,
		RANGAM	,
		RANGE	,
		RANK	,
		RANNOR	,
		RANPOI	,
		RANTBL	,
		RANTRI	,
		RANUNI	,
		RENAME	,
		REPEAT	,
		RESOLVE	,
		REVERSE	,
		REWIND	,
		RIGHT	,
		RMS	,
		ROUND	,
		ROUNDE	,
		ROUNDZ	,
		SASMSG	,
		SASMSGL	,
		SAVING	,
		SAVINGS	,
		SCAN	,
		SDF	,
		SEC	,
		SECOND	,
		SESSBUSY	,
		SETLOCALE	,
		SHA256	,
		SHA256HEX	,
		SHA256HMACHEX	,
		SIGN	,
		SIN	,
		SINH	,
		SKEWNESS	,
		SLEEP	,
		SMALLEST	,
		SOAPWEB	,
		SOAPWEBMETA	,
		SOAPWIPSERVICE	,
		SOAPWIPSRS	,
		SOAPWS	,
		SOAPWSMETA	,
		SORT	,
		SORTKEY	,
		SOUNDEX	,
		SPEDIS	,
		SQLEXEC	,
		SQRT	,
		SQUANTILE	,
		STD	,
		STDDEV	,
		STDERR	,
		STFIPS	,
		STNAME	,
		STNAMEL	,
		STPSRV_HEADER	,
		STPSRV_SESSION	,
		STPSRV_UNQUOTE2	,
		STPSRVGETC	,
		STPSRVGETN	,
		STPSRVSET	,
		STREAMINIT	,
		STRIP	,
		STUDENTS_T	,
		SUBPAD	,
		SUBSTR	,
		SUBSTRING	,
		SUBSTRN	,
		SUM	,
		SUMABS	,
		SYMEXIST	,
		SYMGET	,
		SYMGLOBL	,
		SYMLOCAL	,
		SYSEXIST	,
		SYSGET	,
		SYSMSG	,
		SYSPARM	,
		SYSPROCESSID	,
		SYSPROCESSNAME	,
		SYSPROD	,
		SYSRC	,
		SYSTEM	,
		TAN	,
		TANH	,
		TIME	,
		TIMEPART	,
		TIMEVALUE	,
		TINV	,
		TNONCT	,
		TO_DATE	,
		TO_DOUBLE	,
		TO_TIME	,
		TO_TIMESTAMP	,
		TODAY	,
		TRANSLATE	,
		TRANSTRN	,
		TRANTAB	,
		TRANWRD	,
		TRIGAMMA	,
		TRIM	,
		TRIMN	,
		TRUNC	,
		TYPEOF	,
		TZONEDSTNAME	,
		TZONEDSTOFF	,
		TZONEID	,
		TZONENAME	,
		TZONEOFF	,
		TZONES2U	,
		TZONESTTNAME	,
		TZONESTTOFF	,
		TZONEU2S	,
		UNICODE	,
		UNICODEC	,
		UNICODELEN	,
		UNICODEWIDTH	,
		UNIFORM	,
		UPCASE	,
		URLDECODE	,
		URLENCODE	,
		USS	,
		UUIDGEN	,
		VAR	,
		VARFMT	,
		VARIANCE	,
		VARINFMT	,
		VARLABEL	,
		VARLEN	,
		VARNAME	,
		VARNUM	,
		VARRAY	,
		VARRAYX	,
		VARTRANSCODE	,
		VARTYPE	,
		VERIFY	,
		VFORMAT	,
		VFORMATD	,
		VFORMATDX	,
		VFORMATN	,
		VFORMATNX	,
		VFORMATW	,
		VFORMATWX	,
		VFORMATX	,
		VINARRAY	,
		VINARRAYX	,
		VINFORMAT	,
		VINFORMATD	,
		VINFORMATDX	,
		VINFORMATN	,
		VINFORMATNX	,
		VINFORMATW	,
		VINFORMATWX	,
		VINFORMATX	,
		VLABEL	,
		VLABELX	,
		VLENGTH	,
		VLENGTHX	,
		VNAME	,
		VNAMEX	,
		VTRANSCODE	,
		VTRANSCODEX	,
		VTYPE	,
		VTYPEX	,
		VVALUE	,
		VVALUEX	,
		WAKEUP	,
		WEEK	,
		WEEKDAY	,
		WHICHC	,
		WHICHN	,
		YEAR	,
		YIELDP	,
		YRDIF	,
		YYQ	,
		ZDSATTR	,
		ZDSIDNM	,
		ZDSLIST	,
		ZDSNUM	,
		ZDSRATT	,
		ZDSXATT	,
		ZDSYATT	,
		ZIPCITY	,
		ZIPCITYDISTANCE	,
		ZIPFIPS	,
		ZIPNAME	,
		ZIPNAMEL	,
		ZIPSTATE	,
		ZVOLLIST	,
	},
	morekeywords=[6]{
		_NULL_ ,NULL ,MISSING ,_ALL_ ,_AUTOMATIC_ ,_CHARACTER_ ,_N_ ,_INFILE_
		,_NAME_ ,_NUMERIC_ ,_USER_ ,_WEBOUT_	},
	morekeywords=[7]{ACECLUS	,
		ADAPTIVEREG	,
		ANOM	,
		ANOVA	,
		APPEND	,
		ARIMA	,
		AUTHLIB	,
		AUTOREG	,
		BCHOICE	,
		BGLIMM	,
		BOM	,
		BOXPLOT	,
		CALENDAR	,
		CALIS	,
		CANCORR	,
		CANDISC	,
		CAPABILITY	,
		CATALOG	,
		CATMOD	,
		CAUSALGRAPH	,
		CAUSALMED	,
		CAUSALTRT	,
		CHART	,
		CIMPORT	,
		CLP	,
		CLUSTER	,
		COMPARE	,
		COMPUTAB	,
		CONTENTS	,
		COPULA	,
		COPY	,
		CORR	,
		CORRESP	,
		COUNTREG	,
		CPM	,
		CPORT	,
		CUSUM	,
		DATASETS	,
		DATASOURCE	,
		DATEKEYS	,
		DELETE	,
		DISCRIM	,
		DISPLAY	,
		DISTANCE	,
		DS2	,
		DSTODS2	,
		DTREE	,
		ENTROPY	,
		ESM	,
		EXPAND	,
		EXPORT	,
		FACTEX	,
		FACTOR	,
		FASTCLUS	,
		FCMP	,
		FEDSQL	,
		FMM	,
		FMTC2ITM	,
		FONTREG	,
		FORMAT	,
		FREQ	,
		FREQ	,
		FSLIST	,
		G3D	,
		G3GRID	,
		GA	,
		GAM	,
		GAMPL	,
		GANNO	,
		GANTT	,
		GAREABAR	,
		GBARLINE	,
		GCHART	,
		GCONTOUR	,
		GDEVICE	,
		GEE	,
		GENMOD	,
		GEOCODE	,
		GFONT	,
		GINSIDE	,
		GIS	,
		GKPI	,
		GLIMMIX	,
		GLM	,
		GLMMOD	,
		GLMPOWER	,
		GLMSELECT	,
		GMAP	,
		GOPTIONS	,
		GPLOT	,
		GPROJECT	,
		GRADAR	,
		GREDUCE	,
		GREMOVE	,
		GREPLAY	,
		GROOVY	,
		GSLIDE	,
		GTILE	,
		HADOOP	,
		HDMD	,
		HP4SCORE	,
		HPBIN	,
		HPBNET	,
		HPBOOLRULE	,
		HPCANDISC	,
		HPCDM	,
		HPCDM	,
		HPCLUS	,
		HPCOPULA	,
		HPCOPULA	,
		HPCORR	,
		HPCOUNTREG	,
		HPCOUNTREG	,
		HPDECIDE	,
		HPDMDB	,
		HPDS2	,
		HPF	,
		HPFARIMASPEC	,
		HPFDIAGNOSE	,
		HPFENGINE	,
		HPFESMSPEC	,
		HPFEVENTS	,
		HPFEXMSPEC	,
		HPFIDMSPEC	,
		HPFMM	,
		HPFOREST	,
		HPFRECONCILE	,
		HPFREPOSITORY	,
		HPFSELECT	,
		HPFTEMPRECON	,
		HPFUCMSPEC	,
		HPGENSELECT	,
		HPIMPUTE	,
		HPLMIXED	,
		HPLOGISTIC	,
		HPMIXED	,
		HPNEURAL	,
		HPNLMOD	,
		HPPANEL	,
		HPPANEL	,
		HPPLS	,
		HPPRINCOMP	,
		HPQLIM	,
		HPQLIM	,
		HPQUANTSELECT	,
		HPREDUCE	,
		HPREG	,
		HPSAMPLE	,
		HPSEVERITY	,
		HPSEVERITY	,
		HPSPLIT	,
		HPSUMMARY	,
		HPSVM	,
		HPTMINE	,
		HPTMSCORE	,
		HTTP	,
		ICLIFETEST	,
		ICPHREG	,
		IML	,
		IMPORT	,
		INBREED	,
		INSIGHT ,
		IRT	,
		ISHIKAWA	,
		JAVAINFO	,
		JSON	,
		KDE	,
		KRIGE2D	,
		LATTICE	,
		LIFEREG	,
		LIFETEST	,
		LOAN	,
		LOESS	,
		LOGISTIC	,
		LUA	,
		MACONTROL	,
		MAPIMPORT	,
		MCMC	,
		MDC	,
		MDS	,
		MEANS	,
		MI	,
		MIANALYZE	,
		MIGRATE	,
		MIXED	,
		MODECLUS	,
		MODEL	,
		MULTTEST	,
		MVPDIAGNOSE	,
		MVPMODEL	,
		MVPMONITOR	,
		NESTED	,
		NETDRAW	,
		NLIN	,
		NLMIXED	,
		NPAR1WAY	,
		OPTEX	,
		OPTGRAPH	,
		OPTIONS	,
		OPTLOAD	,
		OPTLP	,
		OPTLSO	,
		OPTMILP	,
		OPTMODEL	,
		OPTNET	,
		OPTQP	,
		OPTSAVE	,
		ORTHOREG	,
		PANEL	,
		PARETO	,
		PDLREG	,
		PHREG	,
		PLAN	,
		PLM	,
		PLOT	,
		PLS	,
		PM	,
		PMENU	,
		POWER	,
		PRESENV	,
		PRINCOMP	,
		PRINQUAL	,
		PRINT	,
		PRINTTO	,
		PROBIT	,
		PRODUCT_STATUS	,
		PROTO	,
		PRTDEF	,
		PRTEXP	,
		PSMATCH	,
		PWENCODE	,
		QDEVICE	,
		QLIM	,
		QUANTLIFE	,
		QUANTREG	,
		QUANTSELECT	,
		RANK	,
		RAREEVENTS	,
		REG	,
		REGISTRY	,
		RELIABILITY	,
		REPORT	,
		RMSTREG	,
		ROBUSTREG	,
		RSREG	,
		S3	,
		SCAPROC	,
		SCORE	,
		SCOREACCEL	,
		SEQDESIGN	,
		SEQTEST	,
		SEVERITY	,
		SGPLOT	,
		SGMAP	,
		SHEWHART	,
		SIM2D	,
		SIMILARITY	,
		SIMLIN	,
		SIMNORMAL	,
		SOAP	,
		SORT	,
		SPATIALREG	,
		SPECTRA	,
		SPP	,
		SQL	,
		SQOOP	,
		SSM	,
		STANDARD	,
		STATESPACE	,
		STDIZE	,
		STDRATE	,
		STEPDISC	,
		STREAM	,
		SUMMARY	,
		SURVEYFREQ	,
		SURVEYIMPUTE	,
		SURVEYLOGISTIC	,
		SURVEYMEANS	,
		SURVEYPHREG	,
		SURVEYREG	,
		SURVEYSELECT	,
		SYSLIN	,
		TABULATE	,
		TIMEDATA	,
		TIMEID	,
		TIMEPLOT	,
		TIMESERIES	,
		TMODEL	,
		TPSPLINE	,
		TRANSPOSE	,
		TRANSREG	,
		TREE	,
		TSCSREG	,
		TTEST	,
		UCM	,
		UNIVARIATE	,
		VARCLUS	,
		VARCOMP	,
		VARIOGRAM	,
		VARMAX	,
		X11	,
		X12	,
		X13	,
		XSL	
	},
	otherkeywords={!,!=,~,\&,_,<,>=,=<,>}
	,
}

\lstalias{sas}{SAS}
\lstdefinestyle{mbfaulmSAS}{
	language          = SAS,
	basicstyle={\normalfont\footnotesize},
	inputencoding=utf8/latin1,
	backgroundcolor=\color{mbfaulmgrissec!10},   
	showstringspaces  = false,   
	showspaces        = false,   
	breaklines        = true,    
	breakatwhitespace = true,    
	columns=fullflexible,	%
	keywordstyle = [2]{\normalfont\bfseries\color{mbfaulmbleumarine}},
	keywordstyle = [3]{\normalfont\color{mbfaulmbleu}},
	keywordstyle = [4]{\normalfont\color{mbfaulmbleumarine}},
	keywordstyle = [5]{\normalfont\color{mbfaulmbleu}},
	keywordstyle = [6]{\normalfont\bfseries\color{mbfaulmbleu}},
	keywordstyle = [7]{\normalfont\bfseries\color{mbfaulmbleumarine}\scshape },
	stringstyle ={\normalfont\color{mbfaulmorangesec}},
	commentstyle = {\normalfont\color{mbfaulmorangesec}},
	literate=
	{á}{{\'a}}1 {é}{{\'e}}1 {í}{{\'i}}1 {ó}{{\'o}}1 {ú}{{\'u}}1
	{À}{{\`A}}1 {à}{{\`a}}1
	{Á}{{\'A}}1 {É}{{\'E}}1 {Í}{{\'I}}1 {Ó}{{\'O}}1 {Ú}{{\'U}}1
	{à}{{\`a}}1 {è}{{\`e}}1 {ì}{{\`i}}1 {ò}{{\`o}}1 {ù}{{\`u}}1
	{À}{{\`A}}1 {È}{{\'E}}1 {Ì}{{\`I}}1 {Ò}{{\`O}}1 {Ù}{{\`U}}1
	{ä}{{\"a}}1 {ë}{{\"e}}1 {ï}{{\"i}}1 {ö}{{\"o}}1 {ü}{{\"u}}1
	{Ä}{{\"A}}1 {Ë}{{\"E}}1 {Ï}{{\"I}}1 {Ö}{{\"O}}1 {Ü}{{\"U}}1
	{â}{{\^a}}1 {ê}{{\^e}}1 {î}{{\^i}}1 {ô}{{\^o}}1 {û}{{\^u}}1
	{Â}{{\^A}}1 {Ê}{{\^E}}1 {Î}{{\^I}}1 {Ô}{{\^O}}1 {Û}{{\^U}}1
	{œ}{{\oe}}1 {Œ}{{\OE}}1 {æ}{{\ae}}1 {Æ}{{\AE}}1 {ß}{{\ss}}1
	{ç}{{\c c}}1 {Ç}{{\c C}}1 {ø}{{\o}}1 {å}{{\r a}}1 {Å}{{\r A}}1
	{€}{{\EUR}}1 {£}{{\pounds}}1
}

\lstdefinestyle{mbfaulmPython}{
	language=Python,
	inputencoding=utf8/latin1,
	backgroundcolor=\color{mbfaulmgrissec!10},   
	commentstyle=\color{mbfaulmorangesec},
	keywordstyle = {\bfseries\color{mbfaulmbleumarine}\scshape \large},
	identifierstyle=\color{mbfaulmbleu},
	stringstyle ={\ttfamily\color{mbfaulmorange}},
	basicstyle={\ttfamily\footnotesize},
	breakatwhitespace=false,         
	breaklines=true,                 
	keepspaces=true,                 
	numbers=left,       
	numbersep=5pt,                  
	showspaces=false,                
	showstringspaces=false,
	showtabs=false,                  
	tabsize=2,
	literate=
	{á}{{\'a}}1 {é}{{\'e}}1 {í}{{\'i}}1 {ó}{{\'o}}1 {ú}{{\'u}}1
	{À}{{\`A}}1 {à}{{\`a}}1
	{Á}{{\'A}}1 {É}{{\'E}}1 {Í}{{\'I}}1 {Ó}{{\'O}}1 {Ú}{{\'U}}1
	{à}{{\`a}}1 {è}{{\`e}}1 {ì}{{\`i}}1 {ò}{{\`o}}1 {ù}{{\`u}}1
	{À}{{\`A}}1 {È}{{\'E}}1 {Ì}{{\`I}}1 {Ò}{{\`O}}1 {Ù}{{\`U}}1
	{ä}{{\"a}}1 {ë}{{\"e}}1 {ï}{{\"i}}1 {ö}{{\"o}}1 {ü}{{\"u}}1
	{Ä}{{\"A}}1 {Ë}{{\"E}}1 {Ï}{{\"I}}1 {Ö}{{\"O}}1 {Ü}{{\"U}}1
	{â}{{\^a}}1 {ê}{{\^e}}1 {î}{{\^i}}1 {ô}{{\^o}}1 {û}{{\^u}}1
	{Â}{{\^A}}1 {Ê}{{\^E}}1 {Î}{{\^I}}1 {Ô}{{\^O}}1 {Û}{{\^U}}1
	{œ}{{\oe}}1 {Œ}{{\OE}}1 {æ}{{\ae}}1 {Æ}{{\AE}}1 {ß}{{\ss}}1
	{ç}{{\c c}}1 {Ç}{{\c C}}1 {ø}{{\o}}1 {å}{{\r a}}1 {Å}{{\r A}}1
	{€}{{\EUR}}1 {£}{{\pounds}}1
}

\lstdefinestyle{mbfaulmVBA}{
	extendedchars=\true,
	language={[Visual]Basic},
	inputencoding=utf8/latin1,
	backgroundcolor=\color{mbfaulmgrissec!10},   
	commentstyle = \color{mbfaulmorangesec}\slshape,
	identifierstyle=\color{mbfaulmbleu},
	keywordstyle = {\color{mbfaulmbleumarine}},
	stringstyle ={\ttfamily\color{mbfaulmorange}},
	basicstyle={\ttfamily\footnotesize},
	breakatwhitespace=false,         
	breaklines=true,                 
	keepspaces=true,                 
	numbers=left,       
	numbersep=5pt,                  
	showspaces=false,                
	showstringspaces=false,
	showtabs=false,                  
	tabsize=2,
	literate=
	{á}{{\'a}}1 {é}{{\'e}}1 {í}{{\'i}}1 {ó}{{\'o}}1 {ú}{{\'u}}1
	{À}{{\`A}}1 {à}{{\`a}}1
	{Á}{{\'A}}1 {É}{{\'E}}1 {Í}{{\'I}}1 {Ó}{{\'O}}1 {Ú}{{\'U}}1
	{à}{{\`a}}1 {è}{{\`e}}1 {ì}{{\`i}}1 {ò}{{\`o}}1 {ù}{{\`u}}1
	{À}{{\`A}}1 {È}{{\'E}}1 {Ì}{{\`I}}1 {Ò}{{\`O}}1 {Ù}{{\`U}}1
	{ä}{{\"a}}1 {ë}{{\"e}}1 {ï}{{\"i}}1 {ö}{{\"o}}1 {ü}{{\"u}}1
	{Ä}{{\"A}}1 {Ë}{{\"E}}1 {Ï}{{\"I}}1 {Ö}{{\"O}}1 {Ü}{{\"U}}1
	{â}{{\^a}}1 {ê}{{\^e}}1 {î}{{\^i}}1 {ô}{{\^o}}1 {û}{{\^u}}1
	{Â}{{\^A}}1 {Ê}{{\^E}}1 {Î}{{\^I}}1 {Ô}{{\^O}}1 {Û}{{\^U}}1
	{œ}{{\oe}}1 {Œ}{{\OE}}1 {æ}{{\ae}}1 {Æ}{{\AE}}1 {ß}{{\ss}}1
	{ç}{{\c c}}1 {Ç}{{\c C}}1 {ø}{{\o}}1 {å}{{\r a}}1 {Å}{{\r A}}1
	{€}{{\EUR}}1 {£}{{\pounds}}1
}

\begin{document}
\title{French wine: Combination of multiple open data sources to mapping the expected harvest value} 
\author{\name{Martial Ph\'elipp\'e-Guinvarc'h\textsuperscript{a} \thanks{\href{mailto:martial.phelippe-guinvarch@univ-lemans.fr}{Martial Ph\'elipp\'e-Guinvarc'h}, Actuary and Lecturer}} \affil{Le Mans University, GAINS, Risques Agricoles SAS. 
	}}
\providecommand{\keywords}[1]{\textbf{\textit{Key words: }} #1}


\maketitle
\begin{keywords}
	open data combination; wine; expected harvest values 
\end{keywords}

\begin{abstract}

	The purpose of this paper is to estimate a representative and detailed map of the harvest value in wine using structured and unstructured open data sources. With climate change and new environmental and ecological policies, wine producers are facing new challenges. 
	The ability to model the evolution of these risks is strategic for wine producers and research in order to adapt.
	Many research projects require the values exposed to risk. For example, to assess the economic impact of risks or the premium of crop insurance, or to choose between different agroecological solutions in a cost-benefit approach. 
	The high spatial heterogeneity and complexity of wine characteristics add to the challenge of these production values and the need to improve our spatial assessment of these harvest-expected values.

	Structured, exhaustive and detailed historical data are collected by the customs services, but they are not open. 
	To achieve this, we combine the aggregate of the vineyard register and the data of the Public Body for Products of Official Quality and Origin. There are several techniques available to merge, combine or complete missing data. We have chosen to use optimization methods to re-estimate the area by appellation and by county, which can then be converted into expected harvest values using olympic average yields by appellation and crop insurance prices. 
	This approach allows us to capture the heterogeneity in production values faced by different vineyards, thereby facilitating further research on risk assessment in the wine industry.

\end{abstract}

\section{Introduction}

The French territory is endowed with a wide variety of lands, appellations, and winemakers' expertise, making each of them unique. 
This diversity represents a significant cultural and gastronomic wealth.
The map of French wine regions and appellations is well-known to wine lovers and experts all over the world and highlights this richness. 
Wine is also a major agricultural product in France and accounted for 15.4\% of all agricultural production with a value of €13.8 billion in 2022 \cite{Nation2022}. In addition, the 2021-2022 campaign saw € 14.28 billion in exports.

With climate change and new environmental and ecological policies and agricultural insurance programs, many impact studies will be carried out.  
For example, the European Commission Farm to Fork Strategy aims to reduce the overall use and risk of chemical pesticides by 50\% and the use of more hazardous pesticides by
50\% by 2030. 
These changes, breaking with the past, drive research on their economic impacts and the evolution of risks.
All of this assessment works require a spatial definition of the value exposed to the risk, and this paper deals with the estimation of a wine crop value map that is as accurate as possible.

Why mapping  the crop wine value ?
First, the heterogeneity of expected values of the harvest in vineyards is greater than in other agricultural productions as cereals due to the presence of appellations.
In viticulture, these appellations designate vineyards with strict regulations on grape growing and wine production practices. These regulations include limits on the maximum yield (measured in hectoliters per hectare, hl/ha) to ensure the quality and distinctiveness of the wine. 
In addition, wines from different appellations can command vastly different prices in the market. The reputation of an appellation, its terroir, and the perceived quality of its wines all contribute to these price differences. As a result, vineyards in prestigious appellations with lower yield limits often produce higher-value wines compared to those in regions with less stringent regulations or lower market recognition.
This combination of diverse yield limits and price disparities leads to a significant variation in the economic value of vineyard harvests.

Second, it is observed that statistics frequently present the value of wine rather than the value of the grape harvest. 
This is due to the fact that these statistics are typically based on sales made by vintners, a significant proportion of whom sell wine in bulk or bottled, rather than selling grapes or must. Only a minimal fraction of the total grape harvest—approximately 8\%—is marketed as fresh grapes, grape juices, or must (\cite{AlonsoUgaglia2019}).
It is, however, important to note that grape production is directly impacted by major climatic and agroecological challenges.
Factors such as temperature changes, precipitation patterns, and extreme weather events can significantly affect the quality and quantity of grapes produced. 
In contrast, the processes of vinification (wine making) and aging are less directly affected by them.
The capacity to store and age wines provides vintners with the option to optimize their product offerings over time, accommodating fluctuations in annual production due to varying climatic and agroecological conditions. 
Even vintage wines, which are often perceived as being composed solely of grapes from a specified year, do not necessarily include 100\% of that year's harvest. 
Regulations typically require that only a minimum of 85\% of the grapes used come from the designated year.
The decoupling of links between harvest values and bottle prices that results from stock and age management is beneficial for winemakers, but presents a challenge for impact studies.

Third, only statistics databases are open, but they do not allow for a detailed mapping of the wine value of production. 
Structured, exhaustive and detailed historical data exist, compiled by customs services.
However,  the confidentiality committee website explains that "statistical confidentiality gives people who provide information used to produce statistics the assurance that this information will not be used in a way that might harm them."

The combination of these three reasons, the  difference between appellations, the importance of knowing the value of the harvest rather than the value of the bottle  wine, and the fact that detailed statistics are not available, makes this work necessary, especially for studies on the economic impact of climate change and on the agroecological transition.   

This paper aims to build and provide a detailed French map of wine values of the harvest using structured and unstructured open data.
This database could also be interpreted as an insurance portfolio for French vineyards, with appellation, location, surface area, quantity, and insurable amount.
We mainly use the vineyard registry statistics and data from The Public Body for Products Under Official Quality and Origin Signs (\href{www.inao.gouv.fr}{INAO}). 
This paper uses constrained minimization optimization methods to combine the data. 

\section{State of the art}

\subsection{Winegrowers face to climate change and  agroecological transition}


The literature reviews by \cite{Ashenfelter2016} and \cite{Wei2022}  demonstrate the dynamic nature of research on the impacts of climate change and the ecological transition in wine. Researchers examine these impacts on the value of wine production, including their components such as price, crop yield, and quality.

The literature on crop yield is quite extensive and diverse, with various studies on crop yield models using growing degree days or hours (\cite{Gu2015}, \cite{Leolini2018}, \cite{Corbeels2018}), on differences between grape varieties, on fighting against frost, heat wave, drought  or hail (\cite{King2016}, \cite{MartinezLuescher2020}, \cite{Ridder2022}, \cite{LopezFornieles2022}), on inputs, and on farming practices in general (\cite{GutierrezGamboa2021}). 
They are often motivated by the need to adapt to climate change and  ecological transition policies.
Of course, to assess the economic relevance of an agronomic solution to the winegrower, the value of the crop is necessary. 
Similarly, to assess the overall impact, the geographic distribution of those values, expected yields  and surfaces will be necessary.

Price is often studied as a time series, as seen in the recent work of \cite{Cardebat2018} on the world wine price index, or in the work of \cite{Paroissien2020} on prices by appellation provided by \textit{Conseil Interprofessionel des Vins de Bordeaux} to forecast prices, or in the work of \cite{CastilloValero2015}, which performs a cointegration analysis on export price series.
Papers explaining wine price heterogeneity typically examine the relationship between price, reputation, quality and age (see the meta-analysis of \cite{Oczkowski2014}%
\footnote{\ One of their results will certainly appeal to wine lovers:
'The most important consequence of the analysis is the relative importance of the notoriety of a wine on its sensory quality, deducing that producers need to maintain the sensory quality of a wine over time to extract Feedback.'}).
\cite{AlonsoUgaglia2019} offer a valuable macro-level analysis of the French wine industry, examining the trends in market prices and production. However, they explain that the fresh grape sales are rare. 
The authors do not feel the need to separate the value of the harvest from the value added by winemaking and aging.

Contrary to expectations, the previously referenced literature reviews by \cite{Ashenfelter2016} and \cite{Wei2022} do not appear to focus on the study of risk, as the term is scarcely employed.
Climate change, in its complexity, encompasses more than a mere rise in average temperature; it also signifies an escalation in extreme weather events. 
Nevertheless, numerous analyses tend to focus solely on average temperatures, neglecting to fully account for extremes.
As \cite{Ashenfelter2016} note, many analyses fail to take into account farmers' potential adaptations, settling for a simplistic “dumb farmer” scenario. 
History has shown that grape growers and winemakers have demonstrated remarkable resilience and adaptability to shifting climatic and economic conditions over thousands of years. 
As such, the evolution of climate risks, the selection of climatic adaptations, and the impact of the agroecological transition will vary from one location to another due to differing climatic conditions and from one appellation to another due to harvest values and winegrowing regulations.

\subsection{Wine mapping}

The literature presents numerous instances of vineyard mapping exercises. Here, we highlight a couple of notable contributions in the field of Precision Viticulture research. \cite{Taylor2018} offers an in-depth study on spatial crop load mapping, while \cite{Gutierrez2021} delves into the intricacies of vineyard water status mapping.

The advent of geo-referenced sensors on harvesting machines has opened up the possibility for high-resolution mapping. This detailed spatial scale allows for a nuanced depiction of heterogeneity (via variogram estimation) within individual parcels, proving invaluable for precision viticulture applications. 
However, the objective of our work is not to achieve such intra-parcel precision. 
Our research primarily targets counties or agricultural regions\footnote{https://agreste.agriculture.gouv.fr/agreste-web/methodon/Z.1/!searchurl/listeTypeMethodon/}, referred to as RAs, which comprise a set of contiguous counties. There are 432 such zones in France.

In a recent proposal, \cite{Bhardwaj2022} generated synthetic data consisting of 1387 samples that mirrored the characteristics of the original data set of 18 samples.
To fit a wine quality prediction model, the authors used the Synthetic Minority Over-Sampling Technique (SMOTE, \cite{Chawla2002}). 
Obtaining a real-world database of comparable size would be prohibitively expensive and challenging.
While the challenge posed by their dataset differs from ours, the recent study under discussion underscores the notion that the construction of representative data is critical to the development of robust models. 
This body of literature will be the focus of the following section of this manuscript.

\subsection{Statistical methods for combining information from multiple data sources}

In this section, we consider statistical methods for combining multiple data sources. 
An essential skill in any programming course is the ability to merge datasets using a common key variable.
This concept, although fundamental, poses a considerable challenge in pedagogy, largely due to the  various merge types such as one-to-one, one-to-many, or many-to-many, and join types such as inner or outer, left join, right join, or full join.\footnote{For example, see the pedagogical source \href{https://guides.nyu.edu/quant/merge}{https://guides.nyu.edu/quant/merge}.} 

The process of merging can indeed be fraught with complications. To illustrate, consider the scenario of merging two datasets on the basis of 'name'. In the first dataset, the format might be 'FirstName Name', while in the second, it could be 'FIRSTNAME NAME'. Similarly, complications may arise when merging datasets by 'zip code' or 'phone number', especially when the first dataset represents these as strings, while the second dataset represents them as numbers with leading zeros and spaces. However, it is reassuring to note that these complexities, while prevalent, are widely recognized, and effective solutions are available across various programming languages.

In contexts such as ours, it is possible that a critical shared variable remains unidentified or turns out to be irrelevant. This shared variable may have been altered or eliminated, or the data sets may represent two different samples from an identical population. In such a scenario, the inner join might fail to produce results even in the presence of the shared variable.

The Panel on Enhancing Federal Statistics for Policy and Social Science Research suggested, \textit{"The amalgamation of data sources and the generation of statistical estimates for desired characteristics can be achieved through the use of record linkage techniques, multiple frame estimation, imputation-based models, and small-area estimation methods"} \cite{FederalStatistics2017a}.

The literature review by \cite{Lohr2017} provides an insightful examination of methods applicable to microdata sets, such as survey data, where each row corresponds to a specific entity, either a legal or natural person. 

Central to these methods is the concept of record linkage, a technique developed to merge records from different data sources that correspond to the same entity. This process can be deterministic, mirroring a straightforward data merge. 
However, an interesting scenario arises when record linkage becomes probabilistic, referred to as PRL. In this case, an algorithm is used to evaluate the similarity of the linkage variables between records from different sources and then retain the most congruent match. 
This process involves the calculation of two probabilities: one, the probability that two records exhibiting this match pattern are actually a correct match, and two, the probability that the identical match pattern could occur by chance between two different entities. 
In the realm of multiple frame estimation methods, various data sources are amalgamated, even in the absence of sufficient identifying information or when survey data spans multiple population sets, with the objective of enhancing the reliability of statistical outcomes.

Imputation-based methods focus primarily on replacing missing data with estimated values. This technique, while seemingly simple, plays a critical role in maintaining data integrity. Consider the example of the Farm Accountancy Data Network longitudinal survey, where initial data on the location of farms within a city are missing. In such cases, the earliest available record for a unit is propagated to all preceding unrecorded data points. Given the low mobility of farmers, this method introduces minimal error. 
Another approach is to replace missing data with an average derived from aggregated data, thus maintaining statistical continuity. 

Alternatively, a predictive model can be used to estimate the missing values. Such a model, when trained on an available subset of the data, provides the ability to simulate or predict a plausible value to fill in the missing data points.
The Multiple Imputation technique, as originally introduced by \cite{Rubin1988}, holds a prominent position in the realm of missing data imputation. 
A nice illustration of its application in economics can be seen in the reproducibility study conducted by \cite{Blundell2008}, where they  generate panel data on consumption for the Panel Study of Income Dynamics (PSID). 
They use an imputation procedure based on food demand estimates from the Consumer Expenditure Survey (CES). 
The model is calibrated using the CES database, taking advantage of the common variables with the PSID database. This model describes a conditional distribution of food consumption, where the dependence is on the common regressors. This conditional distribution is then simulated to serve as the input for the food consumption variable in the PSID database.

Alternative modeling techniques combine individual data points with aggregate statistics to produce estimates for small geographic areas.
This approach is particularly common in agricultural research, where current individual and state-level aggregate crop yields are integrated with historical county statistics to predict current county crop yields. For example, \cite{bellow2007comparison} conducted a comparative study of the Stasny-Goel, Griffith, and simple ratio methods for a variety of crops in ten geographically diverse states, using simulated data sets provided by the National Agricultural Statistics Service, USDA. 
Similarly, the groundbreaking work of \cite{Dombrowski2019,Dombrowski2020} on open data from the National Flood Insurance Program (NFIP) deserves mention. Their study underscores the complexity of merging anonymized databases, such as policy and claim records, from a single vendor. It also highlights the value of merging this data with information from other sources. For example, the researchers integrated NFIP data with American Community Survey (ACS) data to create a visual representation of flood insurance adoption rates.

%

The combination of multiple databases inevitably invites questions about the robustness and potential bias of the estimators used in the models employed. 
A number of studies, for example \cite{Cottrell2009}, establish effectiveness through simulation-based approaches. Conversely, alternative research, exemplified by \cite{Buchinsky2022}, investigates the asymptotic properties of these estimators.

\section{Our model to combine multiple data sources}

\subsection{Used open data}

In this subsection, we elucidate the open data sources that have been utilized to map the wine harvest values. 
While there are additional open data related to the wine industry, these are discussed in Appendix \ref{oso}.

\subsubsection{The vineyard registry statistics}

In the recent advancements of an open data framework, the customs service has made the statistics from the vineyard register accessible (noted CVI). 
It has therefore provided annual statistics (\href{https://www.douane.gouv.fr/la-douane/opendata?f\%5B0\%5D=categorie_opendata_facet\%3A467}{www.douane.gouv.fr}) on area and production, categorized by appellation in one table and by county in another.

The primary strength of this database lies in its reliability. 
The customs services, besides gathering the data, also ensure its dependability. 
As every wine grower is mandated to register their area, production, and stock, the customs service is furnished with comprehensive data. 
However, some statistics are not publicly accessible due to the recommendations set by GDPR (2~127 ha without 'appellation' and 5~936 ha without county). 
In addition, a substantial segment of the Champagne AOP vineyard parcel is excluded from the statistics.

Regrettably, we possess two 'marginal' aggregate tables, but lack the matrix portfolio of values for the harvest, insurable by appellation and county. Given the number of appellations (3~687) and counties (12~645), and in the absence of supplementary information, the quantity of unknown values escalates to nearly 46 million.

\subsubsection{National Institute of origin and quality}


The French Institute for Quality and Origin Product Labels (\href{www.inao.gouv.fr}{INAO}), a public body, offers access to its data via \href{https://www.data.gouv.fr/fr/datasets/data.gouv.fr}{www.data.gouv.fr}. 
Tasked with the recognition and protection of official labels denoting the quality and origin of agricultural, food, and forestry goods, the INAO plays a main role in certifying the geographical origins of wines through appellations. 
Notably, the institute supplies an extensive catalogue of counties (alongside maps in .shp format) where production of listed "Protected Geographical Indication" (IPG) or "Protected Designation of Origin" (AOP) products is authorized. 
Such data is anticipated to significantly enhance our model.

\subsubsection{Other no structured information sources}

It should be noted that the customs data excludes Champagne, a significant component of the French wine industry. Representing 4\% of the nation's total vineyard area, Champagne covers an expansive 34,200~ha as of 2023 and embodies a substantial 27\% of the export value for wines and spirits. More granular data detailing the distribution of vineyard area in each county within this AOP can be found on the \href{https://maisons-champagne.com/fr/filiere/economie/la-champagne-viticole/article/un-acteur-economique-majeur}{maisons-champagne.com} website, a uniquely comprehensive repository of such information. 
This site is key to our initiative to gather information on champagne surfaces, a task impossible to achieve with other resources.

The French Ministry of Agriculture, in the '\emph{specification applicable to insurance companies for the partial assumption of premiums and crop insurance contributions},' establishes the maximum crop scale value allowed to compute subsidized insurance premiums (\href{https://info.agriculture.gouv.fr/boagri/document_administratif-4b9ef75e-29a7-449d-9e40-7e5253bfd642/telechargement}{Available at https://info.agriculture.gouv.fr/}). 
Over time, the level of detail and overall comprehensiveness of the information has increased, and it has become more structured. 
This document states that “the insured price is between 60\% and 120\% of the scale value or, in the absence of reference to the scale, of the actual selling price previously reduced by 17\%”.

Some file manipulation allows us to create structured data that includes wine labels along with their corresponding prices per hectolitre.
Recently, a specific nomenclature of wines was introduced to facilitate subsidies controls work of insurer data \cite{boucher:hal-02481118}. 
In particular, each product is identified as conventional or organic, with the suffix “C” or “B”, respectively.
In the absence of specific information, the conventional price is used.
By choosing to create our own nomenclature rather than adopt the INAO nomenclature, we are faced with the complexities of merging mentioned at the beginning of this section.\footnote{ 
For example, we have used the \texttt{COMPGED} function in SAS, which calculates the distance between two strings to find matches with the INAO nomenclature. 
Manipulations and corrections were sometimes necessary.
Contractions and acronyms were rewritten to be explicit using the \texttt{TRANWRD} function in SAS, such as CDR for Côte du Rhône. 
Conversely, some wine labels combined several appellations, such as "'POMMARD 1er Cru' 'Clos des Epeneaux' 'Les Grands Epenots', 'Les Petits Epenots', 'Les Rugiens Bas', 'Les Rugiens Hauts'". 
The price has therefore been doubled, leaving only one name per row. 
It is also necessary to remove accents and special characters, to standardize the case, and to eliminate meaningless recurring words in the merge such as 'and' or 'of'.
In order to enhance the reliability of the merger, we utilized web scraping techniques on the website, \href{http://www.vin-vigne.com/}{www.vin-vigne.com}. The goal was to incorporate the wine-growing region into the INAO database, a key variable found within the price scale of the insurance specification.
}

Despite the lack of a documented, comprehensive methodology for constructing this scale, it has two advantages. 
The first is that the price source is officially recognized, which has no equivalent, and the second is the level of detail. 
These labels provide scale prices per product, per appellation, per color, or a combination of all, effectively showing the heterogeneity of wine values.

\subsection{Model to combine aggregated data}

Our model aims to estimate a credible surface per county and per product/appellation and converts these surfaces into amount exposed to risk using the listed price  of French Ministry of Agriculture specification and the "expected harvest yields."

The customs services currently provide two distinct open-data sets. The first data set provides statistical information for 8,302 of the 34,839 counties in France, while data for 4,288 other counties remain confidential (the last concerns the harvest of 2022). In the absence of supplementary data, the number of unknown values reaches nearly 9 million, the product of 1,100 appellations and 8,302 counties.
The second data set comprises annual data on non-secretized surfaces and volumes for 2,131 CVI products out of 5,246 wine products, as listed in the INAO database. The last concerns the harvest 2023, while the 2022 data is used to match with the first. The data set includes 2,247 hectares of secretized surfaces, which represent approximately 4.7\% of the total surface area. 
Each CVI product code is a concatenation of wine attributes, including the type of geographical area, wine color, geographical area ID, the kind of wine, and, at least, a product numeric code. In order to facilitate analysis, we have excluded this product numeric code and aggregated by the first part of CVI, which may be referred to for simplicity in the following as appellation. 

Given the limited historical data from 2018 to 2023, the "expected crop yield" is computed using the Olympic average, a method prevalent in multi-peril crop insurance policies. 
The Olympic average can be perceived as a midpoint between the median and the mean, particularly useful when dealing with small datasets. 
It is specified as follows:
\begin{equation}
	\label{MO}
	\tilde{y}^{\mathcal{O}}_{n,a}= \frac{\sum({y}_{n-5\leq k\leq n-1,a} )- \min(y_{n-5\leq k\leq n-1,a}) - \max(y_{n-5\leq k\leq n-1,a}) }{3}
\end{equation}

In this equation, $y_{n,a}$ represents the harvest yields for the year $n$ from appellation $a$.\footnote{In the absence of statistics by appellation, a local Olympic average by type of wine was used. If the value is still missing, a yield of 40 hl/ha has been used for practical reasons.}

\begin{figure}[htp]
\begin{center}
\resizebox{\linewidth}{!}{   \includegraphics{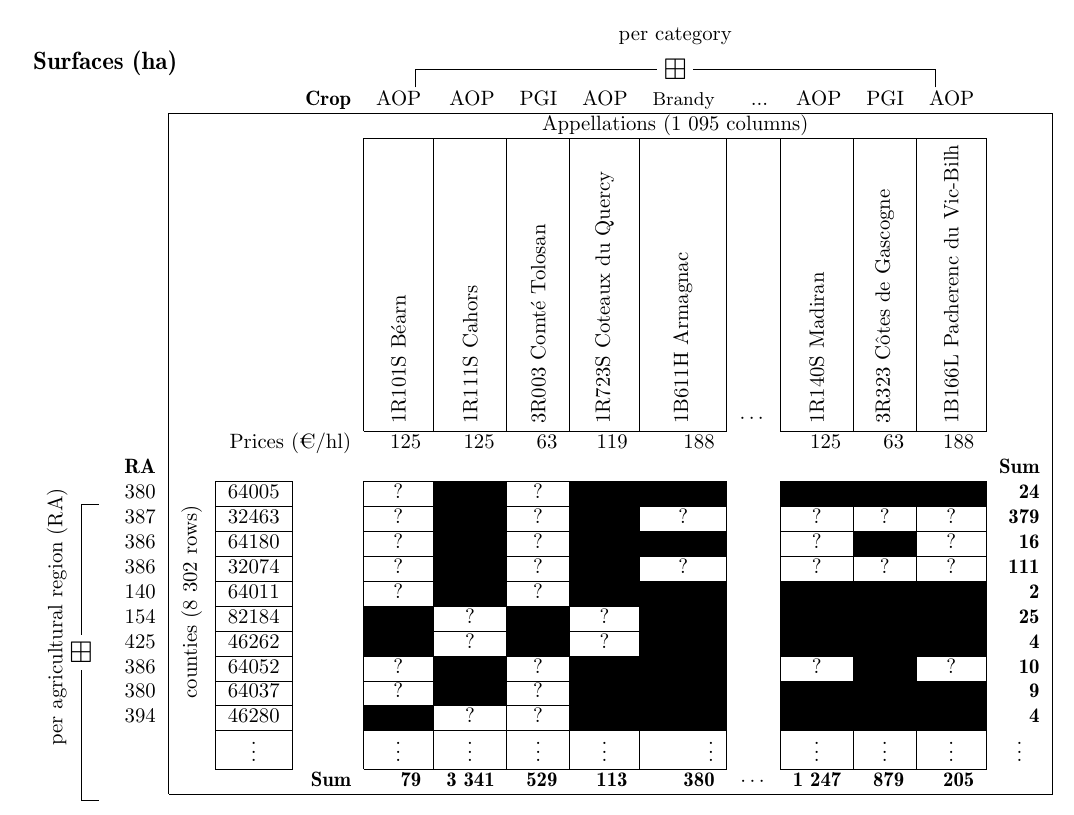}}
\end{center}
\caption{Illustration of the recovering vineyard register data (except Champagne)}
\label{Matrix}
\end{figure}

\href{www.inao.gouv.fr}{INAO} has provided a list of counties where each appellation is authorized. 
As a result, a significant portion of the 9 million unknown values must be set to 0. 
By using the INAO database, only 127,645 surface values are unknown (represented by '?'), significantly reducing the problem. 
On average, an AOP appellation can only be produced in 117 counties, and a winegrower is permitted to produce an average of 16 products/appellations in their county. 
The unknown values can be represented in a matrix with counties in rows and products in columns. 
We have illustrated only a portion of the matrix in Figure \ref{Matrix}. 
The sum by row and column represents the sum of hectare by county and by product/appellation, respectively. 
The black cells indicate unauthorized appellations in a county and have been forced to 0. 
The sign {\Large $\boxplus$} is employed to indicate the possible aggregation of wines into categories or/and agricultural regions (RA).

The aim of this paper is to investigate unknown values through a constrained optimization program as follows:
\begin{equation}
    \begin{array}{lrcll} \displaystyle  {\textrm{maximize}}& \multicolumn{3}{c}{{\displaystyle  \sum _{a\in \mathcal{A}, c \in \mathcal{C}}}\ \alpha_a s_{ac}}  & \\
    \textrm{subject to}& {\displaystyle  \sum _{c \in \mathcal{C}}s_{ac}} & \leq & s_a, & \quad \forall a \in \mathcal{A} \ \ ({Appellations}) \\ 
    & {\displaystyle  \sum _{a \in \mathcal{A}}s_{ac}} & \leq & s_c, & \quad \forall c \in \mathcal{C} \ \ ({Counties}) \\
     & s_{ac} & \geq & 0, & \quad \forall (a,c) \in \mathcal{A} \times \mathcal{C} \\
    \\
    & {\displaystyle  s_{ac} } & =& 0  & \left\lbrace \forall (a,c) \in \mathcal{A} \times \mathcal{C}\right\rbrace\  \bigcap\  \left\lbrace INAO_{ac}=FALSE \right\rbrace  \end{array}
\end{equation}

where $\mathcal{A}$ is the set of appellations, $\mathcal{C}$ is the set of counties, $INAO_{ac} \in \{TRUE,FALSE\}$ is worth $TRUE$ only when  the appellation $a$ is authorized in the county $c$, $s_a$ is the total surface area of appellation $a$, $s_c$ is the total surface area in county $c$, and $ s_{ac}$ is the estimated surface areas of appellation $a$ in county $c$.
The weighted coefficient $\alpha_a$  is designed to prioritize AOP ($\alpha_a=1 $) and PGI wine ($\alpha_a=1/3 $). Non-PGI wines are assigned a coefficient of $\alpha_a=1/4 $. 

One solution to this problem exists. The real surfaces harvested by appellation and by county produce exactly the marginal sums used in the model. 
However, it is counterproductive  to attempt to recover exact information about these surfaces, as this would demonstrate a lack of data secrecy and force customs services to limit data openness. 
There is no presumption that the solution is unique.
The strong conditions imposed on the localization of appellations and the computation of marginal sums enable the generation of plausible estimates for surfaces  by appellation and county.
\footnote{For this process, we employ the SAS proc optmodel, as detailed in the  \href{https://support.sas.com/documentation/onlinedoc/or/143/optmodel.pdf}{SAS PROC OPTMODEL documentation}.} 

The use of random initialization enables the execution of the algorithm from multiple initial points, each with a random uniform value between 0 and the minimum of the two specified values, $S_a$ and $S_c$. This method facilitates the exploration of the solution space in a more effective manner, mitigates the risk of bias by providing unbiased starting points, increases the probability of identifying the global optimum, and avoids the potential for becoming stuck in suboptimal regions.

Non-PGI vineyards are not included in the statistics by appellation. 
In order to have a comprehensive view, a non-PGI pseudo-appellation is created by department. 
Each of these pseudo-appellations is authorized to produce only in each county of the department, and the area of the pseudo-appellation is added to the non-PGI area in the customs statistics.

\section{Results and discussion}
The result of the optimization model is a table of surfaces in hectares per appellation and per county, consisting of 9 million rows, but 746~465 (less than 8.1\%) values are potentially greater than 0. 
The following table~\ref{ExtractOptimization} illustrates this for only three counties in Alsace and five appellations \cite[For the complete result, see]{fwmdataset}. 
The surface area in hectares is the result of the optimization model, the RMO is the historical crop yield Olympic average in hectoliters per hectare from the custom service database, the price in euros per hectoliter is from the insurance specifications, and the crop value is the product of these three previous values. 

It should be noted that even if convergence of optimization is achieved between iterations 36 and 39, the objective value remains unaffected by the initial values.
The Kendall correlation coefficient between the solutions is greater than 99.8\%.
Restricting the county-appellation dataset to solutions with a maximum area of 100 ha or more (1841 rows), the Kendall correlation between solutions remains strong and  greater than 61.49\%. 
The retained values are the average of solutions generated using 20 randomly selected starting values.

The French average of harvest yield is approximately 60.6hl/ha. 
AOPs have a lower mean at 46.5hl/ha in comparison to IPGs at 64.8hl/ha, non-IGP at 60.4hl/ha, and brandy at 121.2hl/ha.
Some wines, such as straw wine, have yields of less than 10 hl/ha, while others, such as brandy and Cremant wines, can have yields of more than 70 hl/ha. These appellation-specific statistics confirm the strong heterogeneity of yields in the wine industry.

\begin{table}[htp] 
	\footnotesize
	\begin{center}
		\begin{tabular}{ll|rrrr}
			\hline
			\textbf{County $^a$} & \textbf{Appellation \& color} & \textbf{Surfaces} & \textbf{Expected Yields} & \textbf{Prices} & \textbf{Harvest Values} \\
			& & (ha) & (hl/ha) & (€/hl) & (€) \\ \hline
			67003 & 1B001M Crémant d'Alsace blanc & 13.5 & 73.16 & 260 & 256,902 \\
			67003 & 1B001S Alsace rouge (Pinot noir) & 19.0 & 64.60 & 260 & 318,670 \\
			67003 & 1B070S Alsace suivi d'un nom de lieu$-$dit blanc  & 3.6 & 44.72 & 260 & 41,939 \\
			67003 & 1R001S Alsace rouge (Pinot noir) & 1.7 & 46.46 & 265 & 20,365 \\
			67003 & 1S001M Crémant d'Alsace rosé & 4.3 & 63.59 & 260 & 70,481 \\
			67003 & 1S001S Alsace rosé (Pinot noir) & 6.6 & 64.36 & 265 & 112,634 \\
			67051 & 1B001M Crémant d'Alsace blanc & 27.5 & 73.16 & 260 & 523,163 \\
			67051 & 1B001S Alsace rouge (Pinot noir) & 67.3 & 64.60 & 260 & 1,130,990 \\
			67051 & 1B053S Alsace grand cru Kaefferkopf & 1.0 & 44.14 & 265 & 12,236 \\
			67051 & 1B070S Alsace suivi d'un nom de lieu$-$dit blanc  & 4.1 & 44.72 & 260 & 48,043 \\
			67051 & 1R001S Alsace rouge (Pinot noir) & 1.8 & 46.46 & 265 & 21,609 \\
			67051 & 1S001M Crémant d'Alsace rosé & 5.0 & 63.59 & 260 & 83,013 \\
			67051 & 1S001S Alsace rosé (Pinot noir) & 8.7 & 64.36 & 265 & 147,972 \\
			67155 & 1B001M Crémant d'Alsace blanc & 28.5 & 73.16 & 260 & 543,080 \\
			67155 & 1B001S Alsace rouge (Pinot noir) & 68.6 & 64.60 & 260 & 1,152,186 \\
			67155 & 1B070S Alsace suivi d'un nom de lieu$-$dit blanc  & 4.2 & 44.72 & 260 & 48,566 \\
			67155 & 1R001S Alsace rouge (Pinot noir) & 1.8 & 46.46 & 265 & 21,707 \\
			67155 & 1S001M Crémant d'Alsace rosé & 5.1 & 63.59 & 260 & 84,120 \\
			67155 & 1S001S Alsace rosé (Pinot noir) & 8.9 & 64.36 & 265 & 151,281 \\
			$\cdots$ & $\cdots$ & $\cdots$ & $\cdots$ & $\cdots$ & $\cdots$\\
			\hline
		\end{tabular}
	\end{center}
	\footnotesize{$^a$ Albé, Blitenschwiller and Gertwiller counties respectively.}
	\caption{Short extract of the optimization model result}
	\label{ExtractOptimization}
\end{table}

The selection of the counties of Albé, Blitenschwiller, and Gertwiller is justified by their red wine production, in contrast to their predominantly white wine-growing region. The table~\ref{ExtractOptimization} reveals that the model is capable of capturing this specificity.

Nevertheless, the model is unable to accurately reflect the diversity observed at the county level. This is evidenced by the fact that changing the starting values does not result in the same solution. Conversely, if the opposite were to be true, it would cast doubt on the confidentiality of the data shared by customs services.

\begin{figure}[htp]
	\centering
	\includegraphics[width=0.7\linewidth]{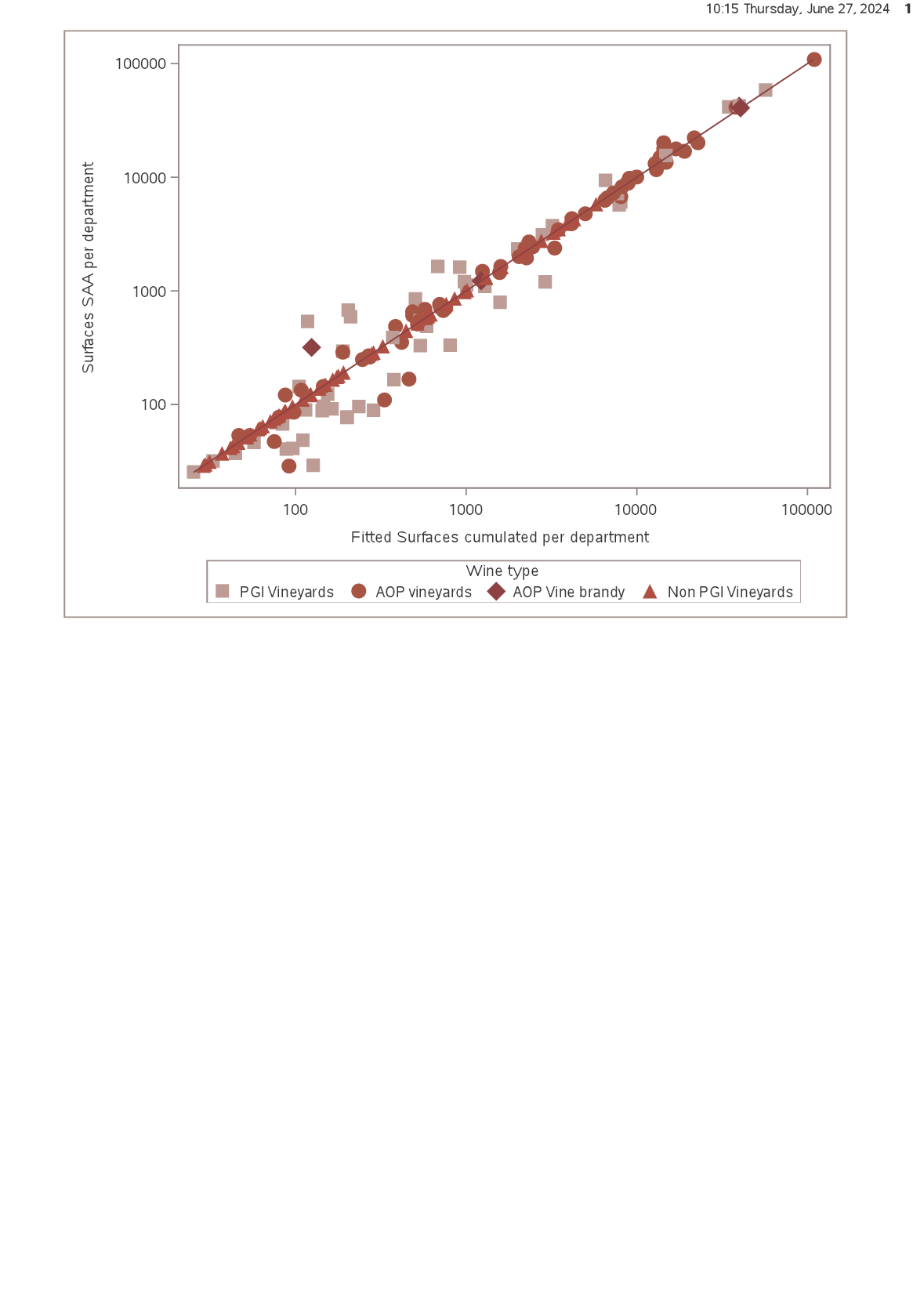}
	\caption{Comparison of aggregated areas with more aggregates statistics}
	\label{fig:comparaisonmodelraa}
\end{figure}

Ideally, we would have compared our results with the undisclosed surface areas of the vineyard register. However, if we aggregate our results by department and wine type, we can compare them with the available statistics in order to obtain a relevant control of our work. We found that the Kendall rate is approximately 88.8\%, and for departments and wine types with a surface area of over 1000 ha, the Kendall rate reaches 89.6\%. 
This result is illustrated in Figure  \ref{fig:comparaisonmodelraa} on a logarithmic scale.
For PGI vineyards, the level of precision is lower due to the very large authorized geographical production areas, which often extend beyond the department scale. 
Conversely, for AOP vineyards, errors are lower due to more restrictive production zones, which are often intra-departmental. 
For  Non-PGI vineyards, the aforementioned errors are reduced at this scale of analysis due to the establishment of pseudo-appellations by department, which compensates for the logical absence of this type of wine in the INAO data set. 
This method of comparison, although less accurate than using the vineyard register, enabled the display of a reasonable level of confidence in the results and the validation of the methodology.

\begin{figure}[htp]
	\centering
	\includegraphics[width=\linewidth]{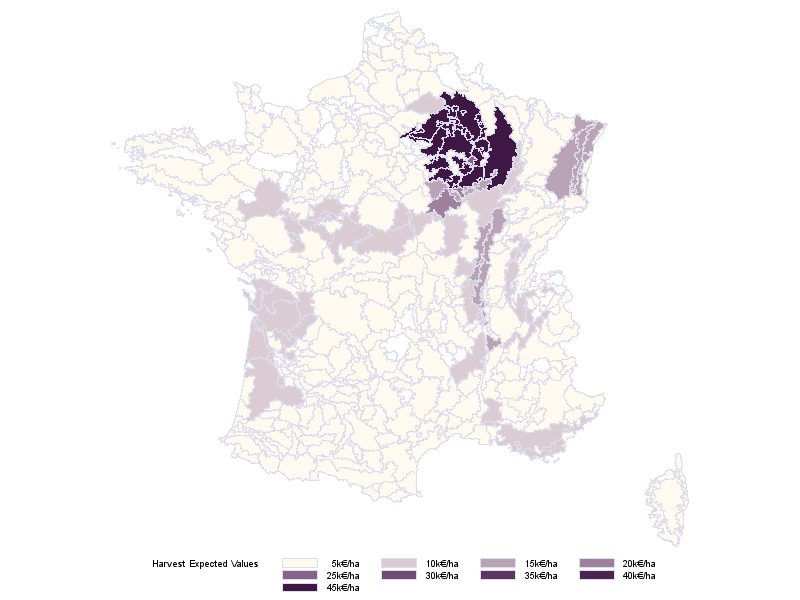}
	\caption{Map of the  harvest values of wine per hectare by agricultural region}
	\label{fig:mapW1}
\end{figure}

\begin{figure}[htp]
	\centering
	\includegraphics[width=\linewidth]{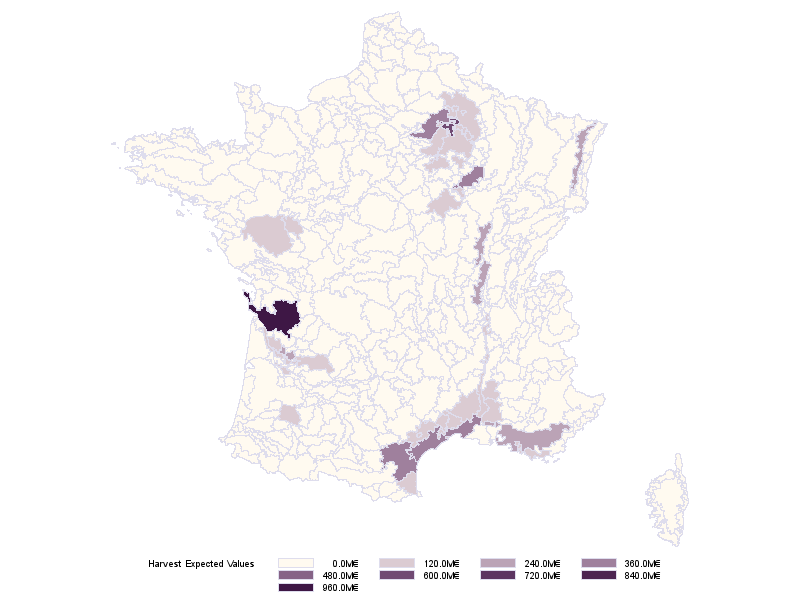}
	\caption{Map of the cumulative harvest values of wine by agricultural region}
	\label{fig:mapW2}
\end{figure}

Figure \ref{fig:mapW1} presents a cartographic representation of the average harvest value per hectare delineated by each agricultural region.
Figure \ref{fig:mapW2} provides a visual compilation of these values. The adoption of this geographical scale, which is approximately four times finer than that of a department, allows for more agronomic coherence and enhanced legibility. 
These harvest value maps can be merged with climate data to facilitate modeling that assesses the economic impact of current and future climate hazards, to test insurance or finance policies or to investigate economically sustainable change in growing practices. 
It should be noted that the financial feasibility and relevance of investments can vary substantially between a crop valued at €5,000 per hectare and one valued at €40,000 per hectare.


Not surprisingly, the highest values per hectare are associated with Champagne. 
However, wines from Alsace have the highest average yield per hectare, suggesting that the region may be more productive than Champagne.
In contrast, wines from Corsica, the Southwest, and Languedoc-Roussillon have lower harvest values.

Comparing these values with the cumulative harvest value completes our interpretation. 
We find agricultural regions that produce Champagne or Alsace wines. However, Cognac, in particular, and Languedoc-Roussillon also emerge as areas with significant sum of harvest values; these large agricultural regions compensate the harvest value per hectare with the cumulative area.

	

The model result, which amounts to €~7.57 billion, can be juxtaposed with the €~13.8 billion reported by the National Institute of Statistics and Economic Studies (INSEE) in  \href{https://agreste.agriculture.gouv.fr/agreste-web/download/publication/publie/Dos2203/2Pages\%20de\%20Dossier2022-3_CCAN_ChapitreII.pdf}{\textit{"Prevision accounting of agriculture"} report}.\footnote{ \href{https://www.insee.fr/fr/statistiques/fichier/6675413/t_10201.xlsx}{Table Insee 2022.} 
A distinctive feature of the data provided by the French National Institute for Statistics and Economic Studies (INSEE) is the presentation of annual production values, segmented by the type of wine. In the context of vineyards, the annual total value of Appellation d'Origine Protégée (AOP), brandy, and other wines (including specific totals for champagne and cognac) is provided.} However, a direct comparison is challenging due to the existing decoupling between the values of the harvest and the prices of the bottles. Insurance specifications present prices based on harvest values, while the INSEE reports values based on vine production. Consequently, the values reported by the INSEE serve solely as a reference point to verify the consistency of the estimated expected values of the harvest. It is reassuring to infer that the overall value of French wine production surpasses the value of the harvest significantly.

\begin{table}[htb]
	\begin{center}
		\begin{tabular}{l|rrrr}
			\hline
			\textbf{} &
			\multicolumn{2}{c}{\textbf{Harvest Expected Value}} &
			\multicolumn{2}{c}{\textbf{Surfaces}} \\ \hline
			\textbf{AOP vineyards}     & 5,208.2M€ & 69.19\% & 429.7Kha & 57.04\% \\
			\textbf{AOP Vine brandy}   & 1,002.3M€  & 13.32\% & 81.8Kha  & 10.86\% \\
			\textbf{PGI Vineyards}     & 1,174.6M€  & 15.60\% & 202.0Kha & 26.81\% \\
			\textbf{Non-PGI Vineyards} & 141.9M€    & 1.89\%  & 39.9Kha  & 5.29\%  \\ \hline
			\multicolumn{1}{c|}{\textbf{Total}} &
			\textbf{7,527.0M€} &
			\textbf{100.00\%} &
			\textbf{753.3Kha} &
			\textbf{100.00\%}
		\end{tabular}
	\end{center}
\end{table}

\section{Conclusions}

Combining data can be challenging. This study highlights the importance of integrating multiple data sources to map harvest values most effectively. The optimization algorithms implemented in the study, combined with the strict constraints imposed by INAO, enabled the production of relevant and useful results. The study led to the creation of an open database, which is highlighted in the article with two maps. The first map shows crop values per hectare, while the second map shows the sum of crop values. 
This allows us to capture the heterogeneity of vineyard characteristics in different regions and to envisage numerous microeconomics studies.

The increase of open data sources represents significant opportunities for research programs dealing with major challenges in the wine industry. However, the implementation of n (\href{https://gdpr-info.eu}{GDPR}) restrictions limits the level of information, particularly the identifying variables, making it difficult to merge multiple relevant data sources to fit models.
This model serves as a further illustration of the potential for overcoming these restrictions and for the efficient combination of multiple data sources.

This study could be enriched in two ways. Firstly, by integrating multi-year data and more recent data or surfaces by counties (not yet available), although this could increase the volume of data during the optimization process and pose difficulties in ensuring consistency between years. 
Secondly, it would be beneficial to consider incorporating production volumes in addition to surface areas as a maximization criterion, although this would increase the complexity of the problem.

Overall, the study provides a valuable demonstration of the potential benefits of combining multiple sources of aggregated data to produce enriched datasets. The results obtained here using this approach have important implications for the agricultural and insurance sectors, where accurate estimates of crop values are essential for efficient decision-making.
	
\section{Acknowledgements}
Thanks to Luc Boucher, Director of DiagoRisk and Risques Agricoles SAS, Jean Cordier, Emerita Professor, and anonymous reviewer for their valuable comments and suggestions.

\bibliography{C:/Users/MARTIA\string~1/GoogleDrive/IRA_MANS/biblio/biblioComplete.bib}
\bibliographystyle{apacite}
\appendix
\section{Wine open sources overview}\label{oso}

This section presents available open data pertinent to the wine industry, accompanied by an analysis of the advantages and limitations of such data.	
\begin{description}

	\item[The FADN open data:]
	
In 2022, this dataset contained information on 1,100 individual accounts of wine growers and survey results on their agronomic practices. 
French viticulture is home to around 59,000 wine-growing professionals, as per the data provided by the 2020 agricultural census (\href{https://agriculture.gouv.fr/infographie-la-viticulture-francaise}{agriculture.gouv.fr}). Although this annual database is highly useful for researchers as it provides various farm-related variables dating back to 2000, it lacks information on the color of wines, varieties, or appellations. In wine, crops are only distinguished as AOP, Vine brandy, PGI Vineyards, Non-PGI Vineyards, and Grapes. Additionally, key variables such as production surface, crop yields, and values were transformed into categorical variables using value ranges for anonymization purposes, and the location variables were removed. 
	
\item[Annual Agricultural Statistics:] 
	
Annual Agricultural Statistics (known as SAA in French) collects annual data on agricultural production, livestock numbers, and areas. Unlike the FADN open database, the SAA data is not anonymized but is aggregated by crop or species, as well as by department. While the FADN and SAA share extensive historical coverage since 1989, the latter does not include more information on wine color, varietals, or appellations.

\item[The graphic land registry database:] 

The graphic land registry database (RPG) has been provided by the service and payment agency (ASP) since 2007 (\href{https://geoservices.ign.fr/documentation/diffusion/telechargement-donnees-libres.html\#rpg}{geoservices.ign.fr}). It is used to calculate public European Common Agricultural Policy (CAP) aid, with greater reliability than a survey. 
It offers an incomparable level of geographical detail on sowing. Even though crop specifications are often more precise than those of the FADN, there are some differences. For example, RPG codes CZH and CZP represent winter and spring rapeseed, respectively, while the FADN code 222 represents both types of rapeseed. In the case of wine, the codes do not distinguish between PGI, AOP, and Non-PGI wines. Although the information provided is very precise and extensive, it only applies to farms that receive CAP aid. However, only a few winegrowers apply for such aid, so the information gathered may not be representative of the entire vineyard plot.

\item[Governmental websites]:

\begin{itemize}
		\item \href{https://www.data.gouv.fr/fr/datasets/agriculture}{www.data.gouv.fr}, 
		\item \href{https://www.insee.fr}{www.insee.fr}, 
		\item \href{https://www.douane.gouv.fr/la-douane/opendata}{www.douane.gouv.fr/la-douane/opendata}, 
		\item 
		\href{https://www.agreste.agriculture.gouv.fr}{www.agreste.agriculture.gouv.fr}
		\item \href{https://geoservices.ign.fr/documentation/donnees/vecteur/rpg}{geoservices.ign.fr}
		\item \href{https://www.geoportail.gouv.fr/donnees/corine-land-cover-2012}{www.geoportail.gouv.fr}
		\item  \href{https://statistiques.msa.fr/geomsa/}{statistiques.msa.fr} 
		\item  \href{https://stats.agriculture.gouv.fr/cartostat/#c=indicator&view=map11}{stats.agriculture.gouv.fr}
	\end{itemize}
	
\item[Other data repertory]:
	
	\begin{itemize}
		\item  \href{https://www.aspexit.com/ou-recuperer-des-sources-de-donnees-en-agriculture}{www.aspexit.com}
		\item   \href{https://www.inao.gouv.fr}{www.inao.gouv.fr},
		\item  \href{https://www.theia-land.fr/product/carte-doccupation-des-sols-de-la-france-metropolitaine/}{www.theia-land.fr}
		\item \href{https://www.liv-ex.com/}{www.liv-ex.com}
		\item \href{https://infoterre.brgm.fr/page/cartes-geologiques}{infoterre.brgm.fr}
		\item \href{https://donneespubliques.meteofrance.fr/}{donneespubliques.meteofrance.fr}, 
		\item \href{https://www.drias-climat.fr/}{www.drias-climat.fr},
		\item  \href{https://agri4cast.jrc.ec.europa.eu/}{agri4cast.jrc.ec.europa.eu}
		\item \href{https://hubeau.eaufrance.fr/}{hubeau.eaufrance.fr}
		\item 
		Each quality sign has specific production rules, such as a list of vine varieties and a cut-off yield. Additionally, each specific production rule for a PGI or an AOP is established through a decree published in the official bulletin of the Ministry of Agriculture and Food. For example, web scraping techniques could be implemented to extract a database of cut-off yields by appellation.
	\end{itemize}
	
\end{description}
At the same time, we are witnessing the enforcement of laws related to privacy protection with regards to personal data. The General Data Protection Regulation (\href{https://gdpr-info.eu}{GDPR}) recommends pseudonymising personal data whenever possible, and anonymising it when necessary. However, this can result in a significant reduction in the richness of the source data.

\section{Code SAS}

The presented an extract of paper SAS code initiates by incorporating datasets retrieved from customs services and INAO. Subsequently, these datasets are combined using the \lstinline[style=mbfaulmSAS]|PROC OPTMODEL| method, as depicted in equation \ref{MO}.

\begin{lstlisting}[style=mbfaulmSAS]
DATA LINAO;
INPUT INSEE_COM CVI AUTHORIZED SINIT;
DATALINES;
01001 3B011 0.33 0
01001 3B011M 0.33 0
01001 3B012 0.33 0
01001 3B012M 0.33 0
01001 3B013 0.33 0
....
;;;;
RUN;

DATA LA;
INPUT  CVI S;
DATALINES;
1B001D 78.714523339
1B001M 3064.7940186
1B001S 9474.2135637
1B002D 3
1B002S 12.987045088
....
;;;;
RUN;

DATA LC;
INPUT INSEE_COM SDC;
DATALINES;
01001 0.2
01002 5.0859
01003 0.2
01004 1.0499999999
01005 0.2
....
;;;;
RUN;

ODS OUTPUT SolutionSummary=SolutionSummary;
PROC OPTMODEL PRESOLVER=AUTOMATIC ;
SET <STR> Icvi;
SET <STR> Icom;
SET <STR,STR> Iinao;
NUM Authorized {Iinao};
NUM SInit {Iinao};
NUM S {Icvi};
NUM SDC {Icom};
READ DATA LA
INTO Icvi=[cvi] S;
READ DATA LC
INTO Icom=[Insee_com] SDC;
READ DATA LINAO 
INTO Iinao=[cvi Insee_com ] Authorized SInit;
SET NODES = union {<cvi,Insee_com> IN Iinao} {cvi,Insee_com};
VAR SurfModel  {<cvi,Insee_com> IN Iinao} INIT SInit[cvi,Insee_com] >= 0 <= MAX(0, MIN(S[cvi],SDC[Insee_com])) ;

MAX obj= SUM {<cvi,Insee_com> IN Iinao} SurfModel[cvi,Insee_com]*Authorized[cvi,Insee_com];
CON SurCVI {cvi IN Icvi}:
SUM {<(cvi),Insee_com> IN Iinao} SurfModel[cvi,Insee_com] <= S[cvi]; 
CON SurCom {Insee_com IN Icom}:
SUM {<cvi,(Insee_com)> IN Iinao} SurfModel[cvi,Insee_com] <= SDC[Insee_com]; 
SOLVE WITH  NLP;
CREATE DATA optmodel(RENAME=(SurfModel=SurfModel&i))
FROM [cvi Insee_com] SurfModel;
QUIT;



\end{lstlisting}

\subsection*{Declaration of AI and AI-assisted technologies in the writing process}
\footnotesize

During the preparation of this work the authors used DeepL and ChatGPT in order to translate French sentence, improve English, convert some tabular or equation to \LaTeX and assist my SAS debugging. After using this tool/service, the authors reviewed and edited the content as needed and take full responsibility for the content of the publication.

\end{document}